\documentclass[pdflatex,sn-mathphys-num]{sn-jnl}

\usepackage{graphicx}%
\usepackage{multirow}%
\usepackage{amsmath,amssymb,amsfonts}%
\usepackage{amsthm}%
\usepackage{mathrsfs}%
\usepackage[title]{appendix}%
\usepackage{xcolor}%
\usepackage{textcomp}%
\usepackage{manyfoot}%
\usepackage{booktabs}%
\usepackage{algorithm}%
\usepackage{algorithmicx}%
\usepackage{algpseudocode}%
\usepackage{listings}%
\usepackage{bm} 
\usepackage{hyperref}
\usepackage[capitalize]{cleveref}
\usepackage{mathtools}
\usepackage{longtable}
\usepackage{pdflscape}

\theoremstyle{thmstyleone}%

\theoremstyle{thmstyletwo}%

\theoremstyle{thmstylethree}%

\begin{document}
	
	\renewcommand{\topfraction}{0.99}
	\renewcommand{\bottomfraction}{0.99}
	\renewcommand{\textfraction}{0.01}
	\renewcommand{\floatpagefraction}{0.8}

	\title[Article Title]{LawMind: A Law-Driven Paradigm for Discovering Analytical Solutions to Partial Differential Equations}
	
	\author[1]{\fnm{Min-Yi} \sur{Zheng}}\email{csmyzheng@mail.scut.edu.cn}
	
	\author*[2]{\fnm{Shengqi} \sur{Zhang}}\email{szhang@eitech.edu.cn}
	
	\author[3]{\fnm{Liancheng} \sur{Wu}}\email{tdwuliancheng@mail.scut.edu.cn}
	
	\author*[1]{\fnm{Jinghui} \sur{Zhong}}\email{jinghuizhong@scut.edu.cn}
	
	\author[2,4]{\fnm{Shiyi} \sur{Chen}}\email{schen@eitech.edu.cn}
	
	\author[5]{\fnm{Yew-Soon} \sur{Ong}}\email{asysong@ntu.edu.sg}
	
	\affil*[1]{\orgdiv{School of Computer Science and Engineering}, \orgname{South China University of Technology}, \orgaddress{\city{Guangzhou}, \postcode{510000}, \country{P.R. China}}}
	
	\affil*[2]{\orgdiv{Zhejiang Key Laboratory of Industrial Intelligence and Digital Twin}, \orgname{Eastern Institute of Technology}, \orgaddress{\city{Ningbo}, \postcode{315200}, \country{P.R. China}}}
	
	\affil[3]{\orgdiv{Department of Tourism Management}, \orgname{South China University of Technology}, \orgaddress{\city{Guangzhou}, \postcode{510000}, \country{P.R. China}}}
	
	\affil[4]{\orgdiv{State Key Laboratory of Turbulence and Complex Systems, College of Engineering}, \orgname{Peking University}, \orgaddress{\city{Beijing}, \postcode{100871}, \country{P.R. China}}}
	
	\affil[5]{\orgdiv{College of Computing and Data Science}, \orgname{Nanyang Technological University}, \orgaddress{\street{Block N4, 2a-28, Nanyang Avenue}, \postcode{639798}, \country{Singapore}}}

	\abstract{Partial differential equations (PDEs) encode fundamental physical laws, yet closed-form analytical solutions for many important equations remain unknown and typically require substantial human insight to derive. Existing numerical, physics-informed, and data-driven approaches approximate solutions from data rather than systematically deriving symbolic expressions directly from governing equations.
	Here we introduce LawMind, a law-driven symbolic discovery framework that autonomously constructs closed-form solutions from PDEs and their associated conditions without relying on data or supervision. By integrating structured symbolic exploration with physics-constrained evaluation, LawMind progressively assembles valid solution components guided solely by governing laws.
	Evaluated on 100 benchmark PDEs drawn from two authoritative handbooks, LawMind successfully recovers closed-form analytical solutions for all cases. Beyond known solutions, LawMind further discovers previously unreported closed-form solutions to both linear and nonlinear PDEs. These findings establish a computational paradigm in which governing equations alone drive autonomous symbolic discovery, enabling the systematic derivation of analytical PDE solutions.}
	
	\maketitle
	
	\section{Introduction}\label{sec:introduction}
	
	Partial differential equations (PDEs) are the mathematical foundation for expressing physical laws that govern diverse natural phenomena. PDEs govern the spatial-temporal evolution of physical fields in domains ranging from fluid mechanics to quantum physics \cite{batchelor2000introduction,shankar2012principles,jackson2012classical,thorne2000gravitation}. Analytical solutions to these equations are fundamental for revealing the mechanisms underlying physical laws. Yet analytical solutions remain scarce, and the difficulty of obtaining them grows rapidly with nonlinearity and equation order. Existing analytical techniques handle only a narrow class of PDEs, leaving most equations without a general framework for deriving exact solutions. Despite advances in computational and data-driven methods, there remains no systematic symbolic framework capable of discovering analytical solutions directly from the governing equations. This gap leaves open whether the physical laws encoded in these equations are sufficient in principle to guide the automated discovery of their own exact solutions.	
	
	A variety of analytical and computational methods have been developed to study PDEs. Analytical methods provide closed-form solutions for a limited set of equations \cite{lie1881integration,olver1993applications,wei1963lie,bluman1989symmetries,ovsiannikov1982group,gardner1967method,ablowitz1974inverse,shabat1972exact,weiss1983painleve,weiss1983painleveII,ablowitz1977exact,liao2003beyond,manas1996darboux,wu2023generalized,wu2022modified,malfliet1992solitary,fan2000extended}. Representative examples include symmetry analysis \cite{lie1881integration,olver1993applications,wei1963lie,bluman1989symmetries,ovsiannikov1982group}, inverse scattering transforms \cite{gardner1967method,ablowitz1974inverse,shabat1972exact}, Painlev\'{e} methods \cite{weiss1983painleve,weiss1983painleveII,ablowitz1977exact} and homotopy techniques \cite{liao2003beyond}. Computational methods have advanced in parallel \cite{courant1928partiellen,crank1947practical,zienkiewicz1965finite,godounov1959difference,orszag1972comparison,leveque2002finite,kutz2016dynamic,lusch2018deep,sirignano2018dgm,raissi2019physics,sun2020surrogate,raissi2020hidden,karniadakis2021physics,mendez2023data,callaham2023multiscale,yin2024scalable,mcgreivy2024weak,brunton2024promising,du2025interpolation}. Classical numerical solvers produce high-accuracy approximations of PDE solutions \cite{courant1928partiellen,crank1947practical,zienkiewicz1965finite,godounov1959difference,orszag1972comparison,leveque2002finite}. Methods such as physics-informed neural networks and operator-learning models obtain numerical solutions from data or from physical constraints \cite{kutz2016dynamic,lusch2018deep,sirignano2018dgm,raissi2019physics,sun2020surrogate,raissi2020hidden,karniadakis2021physics,mendez2023data,callaham2023multiscale,yin2024scalable,mcgreivy2024weak,brunton2024promising,du2025interpolation}.
	These methods leave a gap between symbolic interpretability and computational expressivity. Analytical methods reveal structure but apply only to restricted equation families. Computational methods achieve broad coverage yet yield numerical solutions without symbolic form. To date, no method has treated the governing equation itself as an automated symbolic oracle to systematically search for exact closed-form solutions.
	Since any valid analytical solution must satisfy the governing equation identically, the equation itself provides a natural and self-contained criterion for evaluating candidate symbolic expressions. This observation motivates a law-driven approach in which the governing equation directly guides the search for exact symbolic solutions.
	
	We instantiate this paradigm in LawMind, a symbolic-evolutionary framework that searches the space of candidate symbolic expressions. Each candidate is evaluated by substituting it into the governing PDE to compute a residual-based fitness. This direct substitution requires no external numerical data or collocation sampling, as the governing equation itself imposes selective pressure on the evolving symbolic expressions. To scale this law-driven search to multi-parameter PDEs with complex structure, LawMind employs a hierarchical strategy that decomposes the discovery process into progressive stages, enabling systematic exploration of increasingly complex solution forms. Together, these components establish a law-driven paradigm in which governing equations act as both the problem and the oracle, opening a principled route to analytical discovery in regimes where data are absent or fundamentally unavailable.
	
	We assess LawMind on 100 benchmark PDEs drawn from two authoritative handbooks \cite{polyanin2001handbook,polyanin2003handbook}, spanning five canonical families including first-order PDEs, second-order linear parabolic PDEs, second-order linear hyperbolic PDEs, second-order nonlinear parabolic PDEs, and second-order nonlinear hyperbolic PDEs. LawMind successfully recovers closed-form analytical solutions for all benchmark cases, demonstrating consistent performance across diverse equation structures. Beyond benchmark recovery, we present a case study in which LawMind discovers previously unreported closed-form solutions to a third-order linear PDE and a third-order nonlinear PDE, yielding two new solutions for each equation. This case study illustrates that the law-driven paradigm reveals exact solutions even in regimes where conventional analytic or data-driven methods offer no systematic route. Together, these results demonstrate the law-driven paradigm in action, showing that governing equations themselves can autonomously guide the discovery of exact symbolic solutions and providing a principled framework for diverse PDE problems.
	
	The ability of LawMind to generate exact symbolic solutions directly from governing equations carries broad scientific implications. This capability demonstrates that the governing PDEs themselves serve as autonomous engines of symbolic discovery, providing principled guidance without relying on external numerical data or numerical approximations. This data-free symbolic discovery paradigm contrasts with conventional computational approaches, which rely on training data, numerical approximation, or surrogate models. By placing physical law at the center of analytical exploration, LawMind establishes a route toward fully principled, interpretable solutions that are consistent with the underlying physics. 
	
	\section{Results}\label{sec:results}
	
	\subsection{LawMind overview}
	
	\begin{figure*}[t]
		\centering
		\includegraphics[width=0.9 \textwidth]{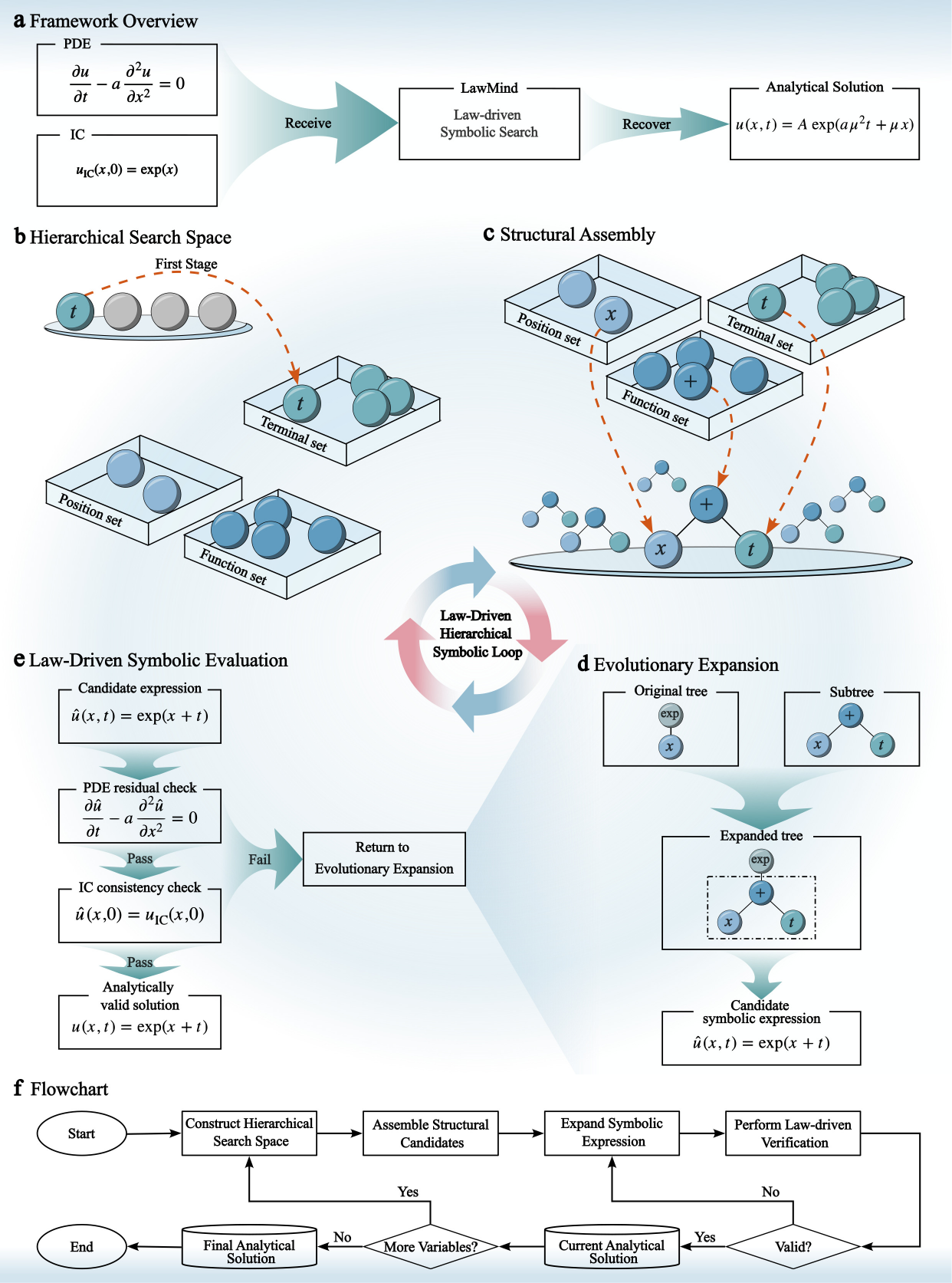}
		\caption{\textbf{LawMind}. 
			\textbf{a}, Given a governing law and associated conditions, LawMind discovers analytical solutions without relying on data or external supervision.
			The example illustrates the recovery of an analytical solution to a heat equation $u_{t} - au_{x,x} = 0$ under prescribed initial conditions $u(x,0) = \exp(x)$.
			\textbf{b-e}, Demonstration of the law-driven symbolic discovery process, including stage-wise activation of variables into the terminal set to incrementally control the symbolic search space (b), assembly of candidate subtrees from these sets to serve as building blocks for expression expansion (c), evolutionary expansion of the current expression by inserting assembled subtrees at designated positions (d), and law-driven evaluation via governing equations and conditions (e).
			\textbf{f}, Overview of the complete LawMind workflow.}
		\label{fig:result1}
	\end{figure*}
	
	PDEs underpin a wide range of physical systems, from diffusion \cite{fourier1822theorie,fick1855liquid} and transport processes \cite{taylor1953dispersion,aris1956dispersion} to fluid dynamics \cite{navier1823memoire,batchelor1967introduction} and kinetic theory \cite{boltzmann1872weitere,cercignani1988boltzmann}. At a mathematical level, they relate an unknown field to its partial derivatives with respect to more than one variable. As an illustrative example, one may consider a field $u(x,t)$ that depends on a spatial coordinate $x$ and a temporal coordinate $t$ and is governed by a nonlinear, parameterized PDE of the form
	$$
	\mathcal{N}[u(x,t);\bm{\lambda}] = 0,
	$$
	where $\mathcal{N}$ denotes a differential operator and $\bm{\lambda}$ represents physical coefficients. In this setting, the governing equation is typically supplemented by initial conditions, while in other cases such as PDEs defined purely over spatial variables, the system may instead involve boundary conditions, source terms, or no additional constraints. More generally, the governing equation together with its associated initial, boundary, or source conditions, collectively denoted as $(\mathcal{N}, C)$, constitute a complete physical specification of the problem. Despite this apparent closure, deriving a closed form analytical solution that exactly satisfies both the PDE and the imposed conditions remains intractable for most nonlinear or high order systems. To address this challenge, we introduce LawMind, a law driven symbolic discovery paradigm that seeks analytical solutions directly from governing laws and their conditions without relying on numerical data or supervised learning.
	
	Given a governing law together with its associated conditions, LawMind performs analytical solution discovery directly from the physical specification of the problem, without relying on numerical data or external supervision (\cref{fig:result1}).
	The illustrative example shown in Fig. \hyperref[fig:result1]{\ref{fig:result1}a} demonstrates the recovery of a closed-form solution to a heat equation subject to prescribed initial conditions.
	
	LawMind follows a hierarchical symbolic discovery paradigm. Fig. \hyperref[fig:result1]{\ref{fig:result1}b-e} depict the law-driven symbolic discovery process carried out within each hierarchical level through adaptive control of the terminal set, structural assembly, evolutionary expansion, and law-driven symbolic evaluation.
	
	To regulate the symbolic search within the hierarchical framework, the terminal set is expanded across successive hierarchical levels (Fig. \hyperref[fig:result1]{\ref{fig:result1}b}). At each level, only a restricted set of variables and constants is made available to the expression, and additional terminals are introduced as the search progresses, enabling systematic growth of analytical structure while controlling the expansion of the symbolic search space.
	
	Given the available terminal set, function set, and position set at a particular hierarchical level, LawMind assembles candidate subtrees by combining elements drawn from these sets (Fig. \hyperref[fig:result1]{\ref{fig:result1}c}). These subtrees serve as structured building blocks for subsequent symbolic expansion, providing a principled starting point for exploration within the current hierarchical level.
	
	Illustrated in Fig. \hyperref[fig:result1]{\ref{fig:result1}d}, the expression is then evolved through insertion of assembled subtrees at designated positions within the current expression, where the expression is initialized from the initial condition or inherited from a previously validated solution. The expansion enables the progressive construction of expressions that remain analytically valid and physically plausible as the search proceeds.
	
	The resulting expression is evaluated in a law-driven manner (Fig. \hyperref[fig:result1]{\ref{fig:result1}e}) through exact symbolic verification against the governing equation and the associated conditions. LawMind performs symbolic computation of the differential operator to determine whether the PDE is satisfied identically rather than approximately or at discrete sample points. If the governing equation is fulfilled, the expression is further checked for compatibility with the prescribed initial condition when available. Based on this verification, fitness is assigned in a strict binary manner, where an expression attains zero fitness if and only if it exactly satisfies both the PDE and the associated conditions, indicating the discovery of an analytical solution. This physics-informed evaluation eliminates the need for training data or numerical simulation and drives the search toward mathematically rigorous solutions rather than numerical approximations.
	
	Each panel in Fig. \hyperref[fig:result1]{\ref{fig:result1}b-\ref{fig:result1}e} illustrates the law-driven symbolic discovery process at the hierarchical level corresponding to the introduction of the variable $t$, demonstrating how the framework operates concretely at a single stage of the progressive search.
	
	Fig. \hyperref[fig:result1]{\ref{fig:result1}f} summarizes the hierarchical and law-driven LawMind workflow, showing how an analytical solution is progressively constructed and rigorously verified through successive cycles of hierarchical search space construction, structural candidate assembly, symbolic expression expansion, and law-driven verification, until a closed-form expression that satisfies the governing equation and associated conditions is identified.
	
	\subsection{Benchmark design and recovery}\label{ss:Benchmark Design and Recovery}
	
	\begin{figure*}[t]
		\centering
		\includegraphics[width=1.0 \textwidth]{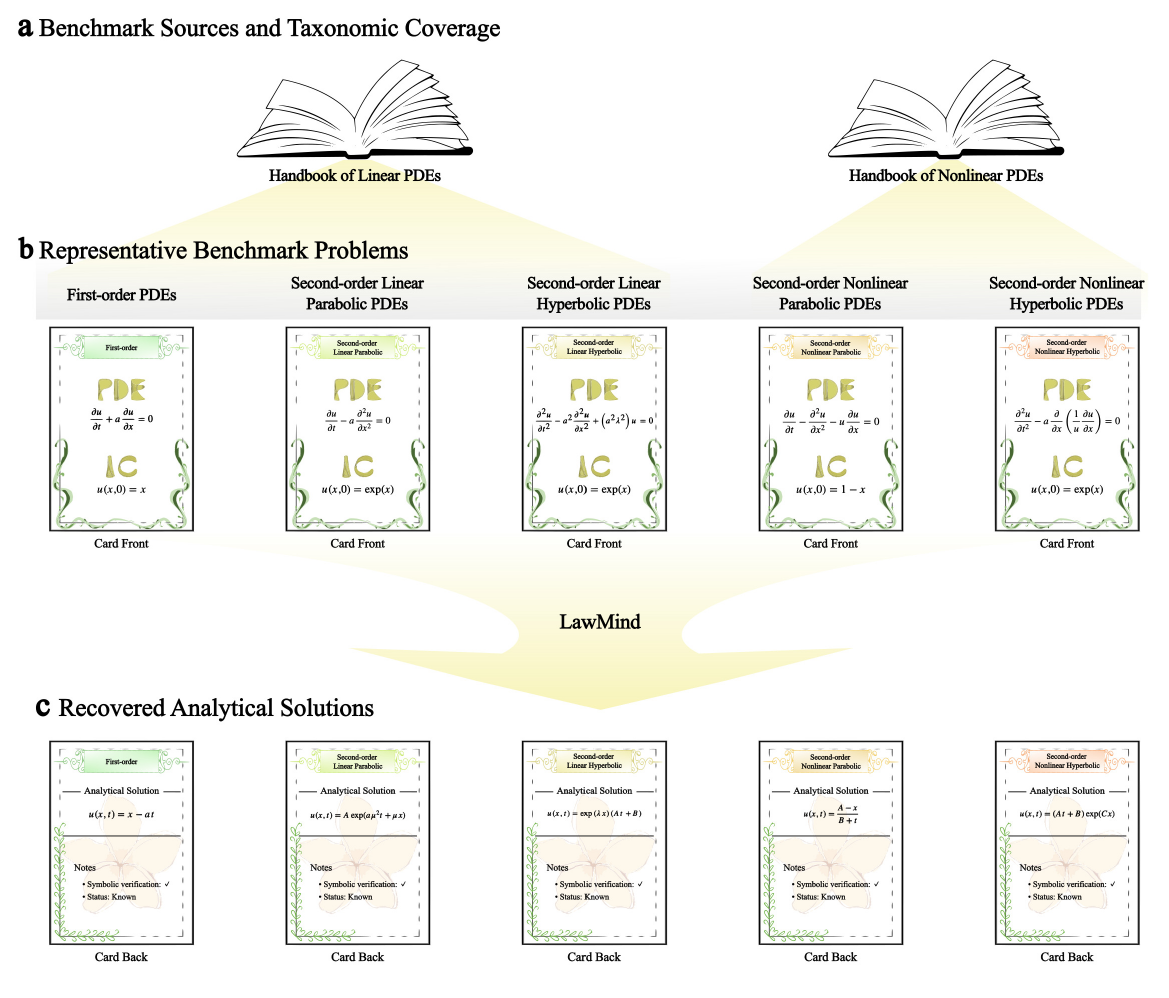}
		\caption{\textbf{Benchmark design and recovery of analytical solutions across five PDE categories}. 
			\textbf{a}, The 100-problem benchmark is curated from two authoritative reference handbooks \cite{polyanin2001handbook,polyanin2003handbook} covering linear and nonlinear PDEs, ensuring broad mathematical coverage. 
			\textbf{b}, Five representative benchmark problems, one from each PDE category: first-order PDEs, second-order linear parabolic PDEs, second-order linear hyperbolic PDEs, second-order nonlinear parabolic PDEs, and second-order nonlinear hyperbolic PDEs. Each problem is presented as a card front specifying the governing PDE and initial condition (IC). 
			\textbf{c}, The corresponding card backs show the closed-form analytical solutions recovered by LawMind. Symbolic verification confirms exact satisfaction of the governing equation and initial condition. Status indicates whether each solution is previously known or newly discovered.}
		\label{fig:result2}
	\end{figure*}
	
	To systematically evaluate the generality of the law-driven symbolic discovery paradigm, we curated a benchmark of 100 PDEs drawn from two authoritative reference handbooks \cite{polyanin2001handbook,polyanin2003handbook} covering linear and nonlinear PDEs. Five representative problems, one from each category, are illustrated in Fig. \hyperref[fig:result1]{\ref{fig:result1}a-b}. The full benchmark is organized across five canonical categories following standard mathematical taxonomy, namely first-order PDEs, second-order linear parabolic PDEs, second-order linear hyperbolic PDEs, second-order nonlinear parabolic PDEs, and second-order nonlinear hyperbolic PDEs, and tabulated in \cref{tab:benchmark_and_result}. Each entry specifies the governing PDE, initial condition, and reference analytical solution in the first three columns.
	
	For each problem, the recovery task requires LawMind to produce a parametric closed-form solution, a symbolic expression retaining free parameters rather than collapsing to a specific numerical instance. A solution is accepted only if it satisfies both the governing PDE and the initial condition under symbolic identity verification, leaving no room for numerical approximation or partial satisfaction.
	
	LawMind successfully recovered closed-form solutions for all 100 benchmark problems, with the recovered solutions documented in the fourth column of \cref{tab:benchmark_and_result}.  Of these, 60 exactly match the parametric forms documented in the reference handbooks, while the remaining 40 are mathematically equivalent parametric representations, as indicated in the fifth column. For instance, for the heat equation family, LawMind recovered $u(x,t) = A+2Bat+Bx^2$, which is mathematically equivalent to the handbook form $u(x,t) = A(x^2+2at)+B$ up to a relabelling of free parameters. Five representative examples are illustrated in Fig. \hyperref[fig:result1]{\ref{fig:result1}c}. The successful verification across this heterogeneous benchmark, encompassing linear and nonlinear equations, parabolic and hyperbolic types, and diverse solution structures, establishes confidence in the correctness and robustness of the law-driven paradigm. This validation on problems with known analytical solutions motivates the application of the law-driven approach to equations where closed-form solutions have not been previously documented.

	\newgeometry{left=1cm, right=1cm, top=2cm, bottom=2cm}
\setlength{\LTcapwidth}{\linewidth}

\begin{landscape}
{\scriptsize
\begin{longtable}{lllll}
\caption{Benchmark PDEs and recovered analytical solutions, grouped by PDE class. $\checkmark$ denotes an exact match with the reference solution. $\equiv$ denotes symbolic equivalence up to algebraic transformation.}
\label{tab:benchmark_and_result} \\
\toprule
{\normalsize\textbf{PDE}} & {\normalsize\textbf{Initial Condition}} & {\normalsize\textbf{Reference Analytical Solution}} & {\normalsize\textbf{Recovered Analytical Solution}} & {\normalsize\textbf{Match}} \\
\midrule
\endfirsthead
\multicolumn{5}{l}{\small\textit{(continued from previous page)}} \\[4pt]
\toprule
{\normalsize\textbf{PDE}} & {\normalsize\textbf{Initial Condition}} & {\normalsize\textbf{Reference Analytical Solution}} & {\normalsize\textbf{Recovered Analytical Solution}} & {\normalsize\textbf{Match}} \\
\midrule
\endhead
\midrule
\multicolumn{5}{r}{\small\textit{(continued on next page)}} \\
\endfoot
\bottomrule
\endlastfoot

\multicolumn{5}{l}{\scriptsize\textbf{First-order PDEs}} \\
\midrule
$\frac{\partial u}{\partial t} + a\frac{\partial u}{\partial x} = 0$ 
& $x$ 
& $x - at$ 
& $- a t + x$ 
& $\checkmark$ \\[6pt]

$a\frac{\partial u}{\partial t} + b\frac{\partial u}{\partial x} = 0$ 
& $\exp(x)$ 
& $\exp(ax - bt)$ 
& $\exp(a x - b t)$ 
& $\checkmark$ \\[6pt]

$(ax+b)\frac{\partial u}{\partial t} + c\frac{\partial u}{\partial x}= 0$ 
& $\sin\left(\frac{x^2}{2}+ x\right)$ 
& $\sin\left(\frac{ax^2}{2} + bx - ct\right)$ 
& $\sin{\left(\frac{a x^{2}}{2} + b x - c t \right)}$ 
& $\checkmark$ \\[6pt]

$a\frac{\partial u}{\partial t} + (bt+c)\frac{\partial u}{\partial x} = 0$ 
& $\cos(x)$ 
& $\cos\left(ax - \frac{bt^2}{2} - ct\right)$ 
& $\cos{\left(- a x + \frac{b t^{2}}{2} + c t \right)}$ 
& $\equiv$ \\[6pt]

$(at-x)\frac{\partial u}{\partial t} + \frac{\partial u}{\partial x}= 0$ 
& $(x+1) \exp(-x)$ 
& $\left(ax-a^2t+1\right)\exp\left(-ax\right)$ 
& $\left(a \left(- a t + x\right) + 1\right) \exp(- a x)$ 
& $\checkmark$ \\[6pt]

$ax\frac{\partial u}{\partial t} + bt\frac{\partial u}{\partial x}= 0$ 
& $\mathrm{erf}\left(x^2\right)$ 
& $\mathrm{erf}\left(ax^2-bt^2\right)$ 
& $\operatorname{erf}{\left(a x^{2} - b t^{2} \right)}$ 
& $\checkmark$ \\[6pt]

$\splitfrac{\left(at^2-cx\right)\frac{\partial u}{\partial t}}{ + t\left(ax+b\right)\frac{\partial u}{\partial x}= 0}$ 
& $\frac{\left(x+1\right)^2}{x^2}$ 
& $\frac{\left(ax+b\right)^2}{cx^2+bt^2}$ 
& $\frac{\left(a x + b\right)^{2}}{b t^{2} + c x^{2}}$ 
& $\checkmark$ \\[12pt]

$\frac{\partial u}{\partial t} + a\exp(\lambda t)\frac{\partial u}{\partial x}= 0$ 
& $x - 1$ 
& $\lambda x - a\exp(\lambda t)$ 
& $- a \exp(\lambda t) + \lambda x$ 
& $\checkmark$ \\[6pt]

$a\exp(\lambda x)\frac{\partial u}{\partial t} + \frac{\partial u}{\partial x}= 0$ 
& $\exp(x)$ 
& $a\exp(\lambda x) - \lambda t$ 
& $a \exp(\lambda x) - \lambda t$ 
& $\checkmark$ \\[6pt]

$\frac{\partial u}{\partial t} + \left(a\exp(\lambda t)+b\right)\frac{\partial u}{\partial x}= 0$ 
& $-x + 1$ 
& $\lambda \left(bt - x\right) + a\exp(\lambda t)$ 
& $\frac{a \exp(\lambda t)}{\lambda} + b t - x$ 
& $\equiv$ \\[6pt]

$\splitfrac{a\exp(-\lambda t)\frac{\partial u}{\partial t}}{ + b\exp(-\beta x)\frac{\partial u}{\partial x}= 0}$ 
& $\exp(x) - 1$ 
& $\frac{1}{\beta b}\exp(\beta x) - \frac{1}{\lambda a}\exp(\lambda t)$ 
& $a \exp(\beta x) - \frac{b \beta \exp(\lambda t)}{\lambda}$ 
& $\equiv$ \\[12pt]

$\splitfrac{\exp(\lambda t) \left(x-a\exp(\mu t)\right)^2\frac{\partial u}{\partial x}}{ +\frac{\partial u}{\partial t} + a\mu \exp(\mu t)\frac{\partial u}{\partial x} =0}$ 
& $\frac{1}{x-1}+1$ 
& $\frac{1}{x-a\exp(\mu t)} + \frac{1}{\lambda}\exp(\lambda t)$ 
& $\frac{\lambda}{- a \exp(\mu t) + x} + \exp(\lambda t)$ 
& $\equiv$ \\[12pt]

$\splitfrac{\left(a\exp(\lambda t) \left(x-b\right)^2\right) \frac{\partial u}{\partial x}}{ +\frac{\partial u}{\partial t} = 0}$ 
& $\frac{1}{x-1}+1$ 
& $\frac{1}{x-b} + \frac{a}{\lambda}\exp(\lambda t)$ 
& $a \exp(\lambda t) + \frac{\lambda}{- b + x}$ 
& $\equiv$ \\[12pt]

$a\exp(\lambda t)\frac{\partial u}{\partial t} + bx \frac{\partial u}{\partial x}= 0$ 
& $\ln(x) + 1$ 
& $\frac{1}{b}\ln(x) + \frac{1}{\lambda a}\exp(-\lambda t)$ 
& $a \ln{\left(x \right)} + \frac{b \exp(- \lambda t)}{\lambda}$ 
& $\equiv$ \\[6pt]

$a\sinh(\lambda x)\frac{\partial u}{\partial t} + \frac{\partial u}{\partial x}= 0$ 
& $-\cosh(x)$ 
& $\lambda t - a\cosh(\lambda x)$ 
& $- a \cosh{\left(\lambda x \right)} + \lambda t$ 
& $\checkmark$ \\[6pt]

$a\cosh(\lambda x)\frac{\partial u}{\partial t} + \frac{\partial u}{\partial x}= 0$ 
& $\sinh(x)$ 
& $a\sinh(\lambda x) - \lambda t$ 
& $a \sinh{\left(\lambda x \right)} - \lambda t$ 
& $\checkmark$ \\[6pt]

$a\tanh(\lambda x)\frac{\partial u}{\partial t} + \frac{\partial u}{\partial x}= 0$ 
& $-\ln\left(\cosh(x)\right)$ 
& $\lambda t - a\ln\left(\cosh(\lambda x)\right)$ 
& $- \ln{\left(\exp(- \frac{\lambda t}{a}) \cosh{\left(\lambda x \right)} \right)}$ 
& $\equiv$ \\[6pt]

$a\frac{\partial u}{\partial t} + b\frac{\partial u}{\partial x} - c= 0$ 
& $\exp(-x)$ 
& $\frac{c}{a}t + \exp(bt-ax)$ 
& $\exp(- a x + b t) + \frac{c t}{a}$ 
& $\checkmark$ \\[6pt]

$a\frac{\partial u}{\partial t} + b\frac{\partial u}{\partial x} - c= 0$ 
& $x + \sin(-x)$ 
& $\frac{c}{b}x + \sin(bt-ax)$ 
& $- \sin{\left(a x - b t \right)} + \frac{c x}{b}$ 
& $\equiv$ \\[6pt]

$a\frac{\partial u}{\partial t} + b\frac{\partial u}{\partial x} - cu = 0$ 
& $\sin(-x)$ 
& $\exp\left(\frac{c}{a}t\right)\sin(bt-ax)$ 
& $- \exp(\frac{c t}{a}) \sin{\left(a x - b t \right)}$ 
& $\equiv$ \\[6pt]

$a\frac{\partial u}{\partial t} + b\frac{\partial u}{\partial x} - cu = 0$ 
& $\exp(x)\cos(-x)$ 
& $\exp\left(\frac{c}{b}x\right)\cos(bt-ax)$ 
& $\exp(\frac{c x}{b}) \cos{\left(a x - b t \right)}$ 
& $\equiv$ \\[6pt]

$\splitfrac{(t-a)\frac{\partial u}{\partial t} + (x-b)\frac{\partial u}{\partial x}}{ - u = 0}$ 
& $(x-1)\exp\left(\frac{-1}{x-1}\right)$ 
& $(x-b)\exp\left(\frac{t-a}{x-b}\right)$ 
& $\left(- b + x\right) \exp(\frac{- a + t}{- b + x})$ 
& $\checkmark$ \\[12pt]

\midrule[0.4pt]
\multicolumn{5}{l}{\scriptsize\textbf{Second-order linear parabolic PDEs}} \\
\midrule
$\frac{\partial u}{\partial t} - a\frac{\partial^2 u}{\partial x^2} = 0$ 
& $x^2$ 
& $x^2+2at$ 
& $2 a t + x^{2}$ 
& $\checkmark$ \\[6pt]

$\frac{\partial u}{\partial t} - a\frac{\partial^2 u}{\partial x^2} = 0$ 
& $x^2+1$ 
& $A(x^2+2at)+B$ 
& $A + 2 B a t + B x^{2}$ 
& $\equiv$ \\[6pt]

$\frac{\partial u}{\partial t} - a\frac{\partial^2 u}{\partial x^2} = 0$ 
& $(x+1)^2$ 
& $A\left(\left(x+B\right)^2+2at\right)$ 
& $2 B a t + B \left(A + x\right)^{2}$ 
& $\equiv$ \\[6pt]

$\frac{\partial u}{\partial t} - a\frac{\partial^2 u}{\partial x^2} = 0$ 
& $\sin (x)$ 
& $A \exp(-a\mu^2 t)\sin \left(\mu x\right)$ 
& $A \exp(- a \mu^2  t) \sin{\left(\mu  x \right)}$ 
& $\checkmark$ \\[6pt]

$\frac{\partial u}{\partial t} - a\frac{\partial^2 u}{\partial x^2} = 0$ 
& $\exp(x)$ 
& $A \exp(a\mu^2 t+\mu x)$ 
& $A \exp(a\mu^2 t+\mu x)$ 
& $\checkmark$ \\[6pt]

$\frac{\partial u}{\partial t} - a\frac{\partial^2 u}{\partial x^2} = 0$ 
& $\frac{1}{2\sqrt{\pi}}\exp\left(-\frac{x^2}{4}\right)$ 
& $\frac{1}{2\sqrt{\pi\left(t+\tau\right)}}\exp\left(-\frac{x^2}{4a\left(t+\tau\right)}\right)$ 
& $\frac{\exp\left(- \frac{x^{2}}{4 a t + 4 \tau}\right)}{2 \sqrt{\pi} \sqrt{a t + \tau}}$ 
& $\checkmark$ \\[10pt]

$\frac{\partial u}{\partial t} - a\frac{\partial^2 u}{\partial x^2} = 0$ 
& $\frac{1}{\sqrt{2}}\exp\left(-\frac{(x-1)^2}{4}\right)$ 
& $\dfrac{1}{\sqrt{2a\left(t+\tau\right)}}\exp\left(-\frac{(x-\mu)^2}{4a\left(t+\tau\right)}\right)$ 
& $\frac{\exp\left(- \frac{\left(- \mu + x\right)^{2}}{4 a t + 4 \tau}\right)}{\sqrt{2 a t + 2 \tau}}$
& $\checkmark$ \\[12pt]

$\frac{\partial u}{\partial t} - a\frac{\partial^2 u}{\partial x^2} = 0$ 
& $\mathrm{erf}\left(\frac{x}{2}\right)$ 
& $\mathrm{erf}\left(\frac{x}{2\sqrt{at+\tau}}\right)$ 
& $\operatorname{erf}{\left(\frac{x}{2 \sqrt{a t + \tau}} \right)}$ 
& $\checkmark$ \\[8pt]

$\frac{\partial u}{\partial t} - a\frac{\partial^2 u}{\partial x^2} - bu = 0$ 
& $x + 1$ 
& $\left(Ax+B\right)\exp\left(bt\right)$ 
& $\left(A + B x\right) \exp(b t)$ 
& $\equiv$ \\[6pt]

$\frac{\partial u}{\partial t} - a\frac{\partial^2 u}{\partial x^2} - bu = 0$ 
& $x^2 + 1$ 
& $\left(A\left(x^2+2at\right)+B\right)\exp\left(bt\right)$ 
& $B \left(A + 2 a t + x^{2}\right) \exp(b t)$ 
& $\equiv$ \\[6pt]

$\frac{\partial u}{\partial t} - a\frac{\partial^2 u}{\partial x^2} - bu = 0$ 
& $\exp(x) + 1$ 
& $A\exp\left(\left(a+b\right)t+ x\right) + B\exp(bt)$ 
& $\left(A + B \exp(a t + x)\right) \exp(b t)$ 
& $\equiv$ \\[6pt]

$\frac{\partial u}{\partial t} - a\frac{\partial^2 u}{\partial x^2} - bu = 0$ 
& $\exp\left(- x\right)\cos(x) + 1$ 
& $\splitfrac{B\exp(bt)}{ + A\exp\left(-x +bt\right)\cos(x-2at)}$ 
& $\splitfrac{\exp(b t)}{ \left(A \exp(- x) \cos{\left(2 a t - x \right)} + B\right)}$ 
& $\equiv$ \\[12pt]

$\frac{\partial u}{\partial t} - a\frac{\partial^2 u}{\partial x^2} - bu = 0$ 
& $\exp\left(- x\right)\sin(x) + 1$ 
& $\splitfrac{B\exp(bt)}{ + A\exp\left(-x +bt\right)\sin(x-2at)}$ 
& $\splitfrac{\exp(b t)}{ \left(- A \exp(- x) \sin{\left(2 a t - x \right)} + B\right)}$ 
& $\equiv$ \\[8pt]

$\frac{\partial u}{\partial t} - a\frac{\partial^2 u}{\partial x^2} - b\frac{\partial u}{\partial x} = 0$ 
& $x$ 
& $x + bt$ 
& $b t + x$ 
& $\checkmark$ \\[6pt]

$\frac{\partial u}{\partial t} - a\frac{\partial^2 u}{\partial x^2} - b\frac{\partial u}{\partial x} = 0$ 
& $x^2$ 
& $\left(x + bt\right)^2 + 2at$ 
& $2 a t + \left(b t + x\right)^{2}$ 
& $\checkmark$ \\[6pt]

$\frac{\partial u}{\partial t} - a\frac{\partial^2 u}{\partial x^2} - b\frac{\partial u}{\partial x} = 0$ 
& $\exp(x)$ 
& $\exp(\frac{2b^2 t + bx}{a})$ 
& $\exp(\frac{b \left(2 b t + x\right)}{a})$ 
& $\checkmark$ \\[6pt]

$\frac{\partial u}{\partial t} - a\frac{\partial^2 u}{\partial x^2} - b\frac{\partial u}{\partial x} = 0$ 
& $\cos(x)$ 
& $\exp\left(-at\right)\cos(x+bt)$ 
& $\exp(- a t) \cos{\left(b t + x \right)}$ 
& $\checkmark$ \\[6pt]

$\frac{\partial u}{\partial t} - a\frac{\partial^2 u}{\partial x^2} - b\frac{\partial u}{\partial x} = 0$ 
& $\sin(x)$ 
& $\exp\left(-at\right)\sin(x+bt)$ 
& $\exp(- a t) \sin{\left(b t + x \right)}$ 
& $\checkmark$ \\[6pt]

$\frac{\partial u}{\partial t} - a\frac{\partial^2 u}{\partial x^2} - b\frac{\partial u}{\partial x} - cu = 0$ 
& $x+1$ 
& $\exp(ct)\left(x + bt + A\right)$ 
& $\left(A + b t + x\right) \exp(c t)$ 
& $\checkmark$ \\[6pt]

$\frac{\partial u}{\partial t} - a\frac{\partial^2 u}{\partial x^2} - (bt+c)\frac{\partial u}{\partial x} = 0$ 
& $2x$ 
& $2x+bt^2+2ct$ 
& $b t^{2} + 2 c t + 2 x$ 
& $\checkmark$ \\[6pt]

$\frac{\partial u}{\partial t} - a\frac{\partial^2 u}{\partial x^2} - bx\frac{\partial u}{\partial x} = 0$ 
& $x$ 
& $x\exp(bt)$ 
& $x \exp(b t)$ 
& $\checkmark$ \\[6pt]

$\frac{\partial u}{\partial t}  - \frac{\partial}{\partial x}\left(\exp(-x)\frac{\partial u}{\partial x}\right) = 0$ 
& $-\exp\left(\frac{x}{2}\right)I_1\left(2\sqrt{i}\exp\left(\frac{x}{2}\right)\right)$ 
& $-\exp\left(\frac{x}{2}+i\omega t\right)I_1\left(2\sqrt{i\omega}\exp\left(\frac{x}{2}\right)\right)$ 
& $- \exp(i \omega t + \frac{x}{2}) I_{1}\left(2 \sqrt{i \omega} \exp(\frac{x}{2})\right)$ 
& $\checkmark$ \\[6pt]

\midrule[0.4pt]
\multicolumn{5}{l}{\scriptsize\textbf{Second-order linear hyperbolic PDEs}} \\
\midrule
$\frac{\partial^2u}{\partial t^2} - a^2\frac{\partial^2u}{\partial x^2} = 0$ 
& $\sin(x) + \exp(x)$ 
& $\sin\left(x+at\right) + \exp\left(x+at\right)$ 
& $\exp(a t + x) + \sin{\left(a t + x \right)}$ 
& $\checkmark$ \\[6pt]

$\frac{\partial^2u}{\partial t^2} - a^2\frac{\partial^2u}{\partial x^2} = 0$ 
& $\cos(x) + \exp(x)$ 
& $\cos\left(\lambda (x+at)\right) + \exp\left(\lambda (x+at)\right)$ 
& $\exp(\lambda \left(a t + x\right)) + \cos{\left(\lambda \left(a t + x\right) \right)}$ 
& $\checkmark$ \\[6pt]

$\frac{\partial^2u}{\partial t^2} - a^2\frac{\partial^2u}{\partial x^2} + a\exp(x) = 0$ 
& $\left(\exp\left(-\frac{x}{4}\right)+\exp\left(\frac{x}{2}\right)\right)^2$ 
& $a\left(A\exp\left(-\frac{t}{4} - \frac{x}{4}\right) + \exp\left(\frac{x}{2}\right)\right)^2$ 
& $a \left(A \exp(- \frac{t}{4} - \frac{x}{4}) + \exp(\frac{x}{2})\right)^{2}$ 
& $\checkmark$ \\[6pt]

$\frac{\partial^2u}{\partial t^2} - a^2\frac{\partial^2u}{\partial x^2} - \mu^2u = 0$ 
& $x+1$ 
& $\exp\left(\mu t\right)\left(Ax+B\right)$ 
& $\left(A + B x\right) \exp(\mu t)$ 
& $\equiv$ \\[6pt]

$\splitfrac{\frac{\partial^2u}{\partial t^2} - a^2\frac{\partial^2u}{\partial x^2}}{ + \left(a^2\lambda^2\right)u = 0}$ 
& $\exp(x)$ 
& $\exp\left(\lambda x\right)\left(At+B\right)$ 
& $\left(A + B t\right) \exp(\lambda x)$ 
& $\equiv$ \\[12pt]

$\splitfrac{\frac{\partial^2u}{\partial t^2} - a^2\frac{\partial^2u}{\partial x^2}}{ + \left(-a^2\lambda^2+\mu^2\right)u = 0}$ 
& $\cos(x)$ 
& $\cos\left(\lambda x + \mu t\right)$ 
& $\cos{\left(\lambda x + \mu t \right)}$ 
& $\checkmark$ \\[12pt]

$\splitfrac{\frac{\partial^2u}{\partial t^2} - a^2\frac{\partial^2u}{\partial x^2}}{ - \left(a^2\lambda^2+\mu^2\right)u = 0}$ 
& $\cos(x) + \sin(x)$ 
& $\exp\left(\mu t\right)\left(A\cos(\lambda x)+B\sin(\lambda x)\right)$ 
& $\left(A \cos{\left(\lambda x \right)} + B \sin{\left(\lambda x \right)}\right) \exp(\mu t)$ 
& $\checkmark$ \\[12pt]

$\frac{\partial^2u}{\partial t^2} - a^2\frac{\partial^2u}{\partial x^2} + bu = 0$ 
& $J_0\left(\sqrt{-x^2}\right)$ 
& $J_0\left(\frac{\sqrt{b}}{a}\sqrt{a^2t^2-x^2}\right)$ 
& $J_{0}\left(\sqrt{b \left(t^{2} - \frac{x^{2}}{a^{2}}\right)}\right)$ 
& $\checkmark$ \\[12pt]

$\frac{\partial^2u}{\partial t^2} - a^2\frac{\partial^2u}{\partial x^2} + bu = 0$ 
& $\splitfrac{J_0\left(\sqrt{1-(x+1)^2}\right)}{+Y_0\left(\sqrt{1-(x+1)^2}\right)}$ 
& $\splitfrac{J_0\left(\frac{\sqrt{b}}{a}\sqrt{a^2(t+A)^2-(x+B)^2}\right)}{+Y_0\left(\frac{\sqrt{b}}{a}\sqrt{a^2(t+A)^2-(x+B)^2}\right)}$ 
& $\splitfrac{J_{0}\left(\sqrt{b} \sqrt{\left(B + t\right)^{2} - \frac{\left(A + x\right)^{2}}{a^{2}}}\right)}{ + Y_{0}\left(\sqrt{b} \sqrt{\left(B + t\right)^{2} - \frac{\left(A + x\right)^{2}}{a^{2}}}\right)}$ 
& $\checkmark$ \\[12pt]

\midrule[0.4pt]
\multicolumn{5}{l}{\scriptsize\textbf{Second-order nonlinear parabolic PDEs}} \\
\midrule
$\frac{\partial u}{\partial t} - \frac{\partial^2 u}{\partial x^2} -u\frac{\partial u}{\partial x} = 0$ 
& $1-x$ 
& $\frac{A-x}{B+t}$ 
& $\frac{B - x}{A + t}$ 
& $\equiv$ \\[6pt]

$\frac{\partial u}{\partial t} - \frac{\partial^2 u}{\partial x^2} -u\frac{\partial u}{\partial x} = 0$ 
& $1+\frac{2}{x+1}$ 
& $1+\frac{2}{x+t+A}$ 
& $1 + \frac{2}{A + t + x}$ 
& $\checkmark$ \\[6pt]

$\frac{\partial u}{\partial t} - \frac{\partial^2 u}{\partial x^2} -u\frac{\partial u}{\partial x} = 0$ 
& $1+\frac{2}{x+1}$ 
& $\lambda+\frac{2}{x+\lambda t+A}$ 
& $\lambda + \frac{2}{A + \lambda t + x}$ 
& $\checkmark$ \\[6pt]

$\frac{\partial u}{\partial t} - \frac{\partial^2 u}{\partial x^2} -u\frac{\partial u}{\partial x} = 0$ 
& $\frac{4x}{x^2+1}$ 
& $\frac{4x}{x^2+2t+A}$ 
& $\frac{4 x}{A + 2 t + x^{2}}$ 
& $\checkmark$ \\[6pt]

$\frac{\partial u}{\partial t} - \frac{\partial^2 u}{\partial x^2} -u\frac{\partial u}{\partial x} = 0$ 
& $\frac{4x+2}{x^2+x}$ 
& $\frac{4x+2A}{x^2+Ax+2t}$ 
& $\frac{2 A + 4 x}{A x + 2 t + x^{2}}$ 
& $\checkmark$ \\[6pt]

$\frac{\partial u}{\partial t} - \frac{\partial^2 u}{\partial x^2} -u\frac{\partial u}{\partial x} = 0$ 
& $\frac{4x+2}{x^2+x+1}$ 
& $\frac{4x+2A}{x^2+Ax+2t+B}$ 
& $\frac{2 B + 4 x}{A + B x + 2 t + x^{2}}$ 
& $\equiv$ \\[6pt]

$\frac{\partial u}{\partial t} - \frac{\partial^2 u}{\partial x^2} -u\frac{\partial u}{\partial x} = 0$ 
& $\frac{2}{1+\exp(-x)}$ 
& $\frac{2}{1+A\exp(-t- x)}$ 
& $\frac{2}{A \exp(- t - x) + 1}$ 
& $\checkmark$ \\[6pt]

$\frac{\partial u}{\partial t} - \frac{\partial^2 u}{\partial x^2} -u\frac{\partial u}{\partial x} = 0$ 
& $1+2\tanh\left(x+1\right)$ 
& $1+2\tanh\left(x+t+A\right)$ 
& $2 \tanh{\left(A + t + x \right)} + 1$ 
& $\checkmark$ \\[6pt]

$\frac{\partial u}{\partial t} - \frac{\partial^2 u}{\partial x^2} -u\frac{\partial u}{\partial x} = 0$ 
& $-1+2\tanh\left(x+1\right)$ 
& $-1 +2\tanh\left(x-t+A\right)$ 
& $2 \tanh{\left(A - t + x \right)} - 1$ 
& $\checkmark$ \\[6pt]

$\frac{\partial u}{\partial t} - \frac{\partial^2 u}{\partial x^2} -u\frac{\partial u}{\partial x} = 0$ 
& $-1+2\tan\left(-x+1\right)$ 
& $-1+2\tan\left(x-t+A\right)$ 
& $2 \tan{\left(A + t - x \right)} - 1$ 
& $\checkmark$ \\[6pt]

$\frac{\partial u}{\partial t} - \frac{\partial^2 u}{\partial x^2} -u\frac{\partial u}{\partial x} = 0$ 
& $\frac{2\cos(x)}{1+\sin(x)}$ 
& $\frac{2\cos(x)}{A\exp(t)+\sin(x)}$ 
& $\frac{2 \cos{\left(x \right)}}{A \exp(t) + \sin{\left(x \right)}}$ 
& $\checkmark$ \\[6pt]

$\frac{\partial u}{\partial t} - \frac{\partial^2 u}{\partial x^2} -u\frac{\partial u}{\partial x} = 0$ 
& $\frac{2\cos(x+1)}{1+\sin(x+1)}$ 
& $\frac{2\cos(x+A)}{B\exp(t)+\sin(x+A)}$ 
& $\frac{2 \cos{\left(B + x \right)}}{A \exp(t) + \sin{\left(B + x \right)}}$ 
& $\equiv$ \\[6pt]

$\frac{\partial u}{\partial t} - \frac{\partial^2 u}{\partial x^2} -u\frac{\partial u}{\partial x} = 0$ 
& $\frac{2\cos\left(x + 1\right)}{1 + \sin(x+1)}$ 
& $\frac{2\cos\left(x + A\right)}{B\exp(t) + \sin(x+A)}$ 
& $\frac{2 \cos{\left(B + x \right)}}{A \exp(t) + \sin{\left(B + x \right)}}$ 
& $\equiv$ \\[6pt]

$\frac{\partial u}{\partial t} - \frac{\partial^2 u}{\partial x^2} -u\frac{\partial u}{\partial x} = 0$ 
& $-1+\frac{\exp(x)-1}{\exp(x)+1}$ 
& $\frac{-A + B\exp(-t + x)}{A + B\exp(-t + x)} - 1$ 
& $\frac{- A + B \exp(- t + x)}{A + B \exp(- t + x)} - 1$ 
& $\checkmark$ \\[6pt]

$\frac{\partial u}{\partial t} - a\frac{\partial^2 u}{\partial x^2} -bu\frac{\partial u}{\partial x} = 0$ 
& $\frac{1}{\exp(t - x)}$ 
& $\frac{1}{A \exp(-a t - x) + \frac{b}{2a}}$ 
& $\frac{1}{A \exp(- a t - x) + \frac{b}{2 a}}$ 
& $\checkmark$ \\[6pt]

$\frac{\partial u}{\partial t} - a \frac{\partial^2 u}{\partial x^2} - b\left(\frac{\partial u}{\partial x}\right)^2 = 0$ 
& $x+1$ 
& $Ax+A^2bt+B$ 
& $A + B x + B^2 b t$ 
& $\equiv$ \\[6pt]

$\frac{\partial u}{\partial t} - a u \frac{\partial^2 u}{\partial x^2} = 0$ 
& $x^2+x+1$ 
& $\frac{x^2+Ax+B}{C-2at}$ 
& $\frac{A x + C + x^{2}}{B - 2 a t}$ 
& $\equiv$ \\[6pt]

$\frac{\partial u}{\partial t} - a u \frac{\partial^2 u}{\partial x^2} -b = 0$ 
& $x+1$ 
& $Ax+bt+B$ 
& $A x + B + b t$ 
& $\checkmark$ \\[6pt]

$\frac{\partial u}{\partial t} - a u \frac{\partial^2 u}{\partial x^2} - bu - c = 0$ 
& $x - 1$ 
& $Ax\exp(bt)-\frac{c}{b}$ 
& $A x \exp(b t) - \frac{c}{b}$ 
& $\checkmark$ \\[6pt]

$\frac{\partial u}{\partial t} - a \frac{\partial}{\partial x}\left(\frac{1}{u}\frac{\partial u}{\partial x}\right) = 0$ 
& $\frac{1}{x+1}$ 
& $\frac{1}{Ax-aA^2t+B}$ 
& $\frac{1}{A + B x - B^2 a t}$ 
& $\equiv$ \\[6pt]

$\frac{\partial u}{\partial t} - a \frac{\partial}{\partial x}\left(\frac{1}{u}\frac{\partial u}{\partial x}\right) = 0$ 
& $\frac{1}{(x+1)^2}$ 
& $\frac{2at+A}{\left(x+B\right)^2}$ 
& $\frac{B + 2 a t}{\left(A + x\right)^{2}}$ 
& $\equiv$ \\[6pt]

$\frac{\partial u}{\partial t} - a \frac{\partial}{\partial x}\left(\frac{1}{u}\frac{\partial u}{\partial x}\right) = 0$ 
& $\frac{1}{1+\exp(-x)}$ 
& $\frac{1}{1+A\exp(at-x)}$ 
& $\frac{1}{A \exp(a t - x) + 1}$ 
& $\checkmark$ \\[6pt]

$\frac{\partial u}{\partial t} - a \frac{\partial}{\partial x}\left(\frac{1}{u}\frac{\partial u}{\partial x}\right) = 0$ 
& $\frac{1}{\sinh^2(x+1)}$ 
& $\frac{2at+A}{\sinh^2(x+B)}$ 
& $\frac{B + 2 a t}{\sinh^{2}{\left(A + x \right)}}$ 
& $\equiv$ \\[6pt]

$\frac{\partial u}{\partial t} - a \frac{\partial}{\partial x}\left(\frac{1}{u}\frac{\partial u}{\partial x}\right) = 0$ 
& $\frac{1}{\cosh^2(x+1)}$ 
& $\frac{A-2at}{\cosh^2(x+B)}$ 
& $\frac{B - 2 a t}{\cosh^{2}{\left(A + x \right)}}$ 
& $\equiv$ \\[6pt]

$\splitfrac{\frac{\partial u}{\partial t} - \frac{\partial}{\partial x}\left(\exp(u)\frac{\partial u}{\partial x}\right)}{ - \exp(u) - a = 0}$ 
& $\ln\left(\cos(x)+\sin(x)\right)$ 
& $\ln\left(A\cos(x)+B\sin(x)\right)+ at$ 
& $a t + \ln{\left(A \cos{\left(x \right)} + B \sin{\left(x \right)} \right)}$ 
& $\checkmark$ \\[12pt]

\midrule[0.4pt]
\multicolumn{5}{l}{\scriptsize\textbf{Second-order nonlinear hyperbolic PDEs}} \\
\midrule
$\frac{\partial^2 u}{\partial t^2} - \frac{\partial^2 u}{\partial x^2} - au^2 = 0$ 
& $\frac{18}{x^2}$ 
& $\frac{18}{a\left(x+2t\right)^2}$ 
& $\frac{18}{a \left(2 t + x\right)^{2}}$ 
& $\checkmark$ \\[6pt]

$\frac{\partial^2 u}{\partial t^2} - \frac{\partial^2 u}{\partial x^2} - au^2 = 0$ 
& $\frac{4}{1-\left(x+1\right)^2}$ 
& $\frac{4}{a\left(\left(t+A\right)^2-\left(x+B\right)^2\right)}$ 
& $\frac{4}{- a \left(A + x\right)^{2} + a \left(B + t\right)^{2}}$ 
& $\equiv$ \\[6pt]

$\frac{\partial^2 u}{\partial t^2} - \frac{\partial^2 u}{\partial x^2} - a\exp(x)u^2 = 0$ 
& $\left(\exp\left(\frac{x}{12}\right)+\exp\left(\frac{x}{2}\right)\right)^{-2}$ 
& $a\left(A\exp\left(\frac{t}{12} + \frac{x}{12}\right) + a\exp\left(\frac{x}{2}\right)\right)^{-2}$ 
& $\frac{a}{\left(A \exp(\frac{t}{12} + \frac{x}{12}) + a \exp(\frac{x}{2})\right)^{2}}$ 
& $\checkmark$ \\[6pt]

$\frac{\partial^2 u}{\partial t^2} - a\left(\frac{\partial^2 u}{\partial x^2}+\frac{1}{x}\frac{\partial u}{\partial x}\right) - \frac{b}{u} = 0$ 
& $\sqrt{\frac{1}{2}\left(1-x^2\right)}$ 
& $\sqrt{\frac{b}{2a}\left(a\left(t+A\right)^2-x^2\right)}$ 
& $\sqrt{\frac{b \left(A + t\right)^{2}}{2} - \frac{b x^{2}}{2 a}}$ 
& $\checkmark$ \\[6pt]

$\frac{\partial^2 u}{\partial t^2} - au\frac{\partial^2 u}{\partial x^2} = 0$ 
& $\frac{3(x+1)^2}{a}$ 
& $\frac{3(x+A)^2}{a(t+B)^2}$ 
& $\frac{3 \left(B + x\right)^{2}}{a \left(A + t\right)^{2}}$ 
& $\equiv$ \\[6pt]

$\frac{\partial^2 u}{\partial t^2} - a\frac{\partial }{\partial x}\left(u\frac{\partial u}{\partial x}\right) = 0$ 
& $\left(x+1\right)^2$ 
& $\frac{1}{a}\left(\frac{x+A}{t+B}\right)^2$ 
& $\frac{\left(B + x\right)^{2}}{a \left(A + t\right)^{2}}$ 
& $\equiv$ \\[6pt]

$\frac{\partial^2 u}{\partial t^2} - a\frac{\partial }{\partial x}\left(u\frac{\partial u}{\partial x}\right) = 0$ 
& $\sqrt{x+1}$ 
& $\left(At+B\right)\sqrt{Cx+D}$ 
& $\sqrt{A + B x} \left(C + D t\right)$ 
& $\equiv$ \\[6pt]

$\frac{\partial^2 u}{\partial t^2} - a\frac{\partial }{\partial x}\left(\frac{1}{u}\frac{\partial u}{\partial x}\right) = 0$ 
& $\exp(x)$ 
& $\left(At+B\right)\exp(Cx)$ 
& $\left(B + C t\right) \exp(A x)$ 
& $\equiv$ \\[6pt]

$\frac{\partial^2 u}{\partial t^2} - a\frac{\partial }{\partial x}\left(\frac{1}{u}\frac{\partial u}{\partial x}\right) = 0$ 
& $\frac{1}{\left(x+1\right)^2}$ 
& $\frac{a t^{2} + At + B}{(x+C)^{2}}$ 
& $\frac{B t + C + a t^{2}}{\left(A + x\right)^{2}}$ 
& $\equiv$ \\[6pt]

$\frac{\partial^2 u}{\partial t^2} - a\frac{\partial }{\partial x}\left(\frac{1}{u}\frac{\partial u}{\partial x}\right) = 0$ 
& $\frac{1}{\cosh^2\left(x+1\right)}$ 
& $\frac{- a t^{2} - t + 1}{\cosh^{2}{\left(x + 1\right)}}$ 
& $\frac{- a t^{2} - t + 1}{\cosh^{2}{\left(x + 1 \right)}}$ 
& $\checkmark$ \\[6pt]

$\frac{\partial^2 u}{\partial t^2} - a\frac{\partial }{\partial x}\left(\frac{1}{u}\frac{\partial u}{\partial x}\right) = 0$ 
& $\frac{1}{\sinh^2\left(x+1\right)}$ 
& $\frac{a t^{2} + t + 1}{\sinh^{2}{\left(x + 1\right)}}$ 
& $\frac{a t^{2} + t + 1}{\sinh^{2}{\left(x + 1 \right)}}$ 
& $\checkmark$ \\[6pt]

$\frac{\partial^2 u}{\partial t^2} - a\frac{\partial }{\partial x}\left(\frac{1}{u}\frac{\partial u}{\partial x}\right) = 0$ 
& $\frac{1}{\cos^2\left(x+1\right)}$ 
& $\frac{a t^{2} + t + 1}{\cos^{2}{\left(x+1 \right)}}$ 
& $\frac{a t^{2} + t + 1}{\cos^{2}{\left(x + 1 \right)}}$ 
& $\checkmark$ \\[6pt]

$\frac{\partial^2 u}{\partial x^2} - \frac{1}{a}\frac{\partial }{\partial t}\left(u\frac{\partial u}{\partial t}\right) = 0$ 
& $\left(\frac{1}{x+1}\right)^2$ 
& $a\left(\frac{t+A}{x+B}\right)^2$ 
& $\frac{a \left(A + t\right)^{2}}{\left(B + x\right)^{2}}$ 
& $\checkmark$ \\[6pt]

$\frac{\partial^2 u}{\partial x^2} - \frac{1}{a}\frac{\partial }{\partial t}\left(u\frac{\partial u}{\partial t}\right) = 0$ 
& $x+1$ 
& $\left(Ax+B\right)\sqrt{Ct+D}$ 
& $\sqrt{A + B t} \left(C + D x\right)$ 
& $\equiv$ \\[6pt]

$\frac{\partial^2 u}{\partial x^2} - \frac{1}{a}\frac{\partial }{\partial t}\left(u\frac{\partial u}{\partial t}\right) = 0$ 
& $\sqrt{x+1} + 1$ 
& $B \sqrt{t+x+A} + a$ 
& $B \sqrt{A + t + x} + a$ 
& $\checkmark$ \\[6pt]

$\frac{\partial^2 u}{\partial t^2} - a^2\frac{\partial^2 u}{\partial x^2} - b\exp(u) = 0$ 
& $-2\ln\left(\exp(x)+\frac{\sqrt{2}}{2}\sinh(1)\right)$ 
& $-2\ln\left(A\exp(x)+\frac{\sqrt{2b}}{2a}\sinh(at+B)\right)$ 
& $- 2 \ln{\left(B \exp(x) + \frac{\sqrt{2} \sqrt{b} \sinh{\left(A + a t \right)}}{2 a} \right)}$ 
& $\equiv$ \\[6pt]

$\frac{\partial^2 u}{\partial t^2} - a^2\frac{\partial^2 u}{\partial x^2} - b\exp(u) = 0$ 
& $-2\ln\left(1+\frac{\sqrt{2}}{2}\cosh(x+1)\right)$ 
& $-2\ln\left(A\exp(at)+\frac{\sqrt{2b}}{2a}\cosh(x+B)\right)$ 
& $- 2 \ln{\left(B \exp(a t) + \frac{\sqrt{2} \sqrt{b} \cosh{\left(A + x \right)}}{2 a} \right)}$ 
& $\equiv$ \\[6pt]

$\splitfrac{\frac{\partial^2 u}{\partial t^2} - \frac{a}{x}\frac{\partial}{\partial x}\left(x\frac{\partial u}{\partial x}\right)}{ - b\exp(u) = 0}$ 
& $-\ln\left(\frac{1}{2}\left(x^2-1\right)\right)$ 
& $-\ln\left(\frac{b}{2a}\left(x^2-a(t+A)^2\right)\right)$ 
& $- \ln{\left(- \frac{b \left(A + t\right)^{2}}{2} + \frac{b x^{2}}{2 a} \right)}$ 
& $\checkmark$ \\[12pt]

$\splitfrac{\frac{\partial^2 u}{\partial t^2} - \frac{\partial}{\partial x}\left(ax\frac{\partial u}{\partial x}\right)}{ - b\exp(\lambda u) = 0}$ 
& $-\ln\left(2\left(x-\frac{1}{4}\right)\right)$ 
& $-\frac{1}{\lambda}\ln\left(2b\lambda\left(\frac{x}{a}-\frac{\left(t+A\right)^2}{4}\right)\right)$ 
& $- \frac{\ln{\left(\lambda \left(- \frac{b \left(A + t\right)^{2}}{2} + \frac{2 b x}{a}\right) \right)}}{\lambda}$ 
& $\checkmark$ \\[12pt]

$\frac{\partial^2 u}{\partial t^2} - a\exp(\lambda u)\frac{\partial^2 u}{\partial x^2} = 0$ 
& $\ln\left(\cosh^2(x+1)\right)$ 
& $\frac{1}{\lambda}\ln\left(\frac{\cosh^2(x+A)}{a\cos^2(t)}\right)$ 
& $\frac{\ln{\left(\frac{\cosh^{2}{\left(A + x \right)}}{a \cos^{2}{\left(t \right)}} \right)}}{\lambda}$ 
& $\checkmark$ \\[6pt]

$\frac{\partial^2 u}{\partial t^2} - a\exp(\lambda u)\frac{\partial^2 u}{\partial x^2} = 0$ 
& $\ln\left(\frac{\left(x+1\right)^2}{\cosh^2(1)}\right)$ 
& $\frac{1}{\lambda}\ln\left(\frac{1}{a}\frac{\left(x+A\right)^2}{\cosh^2(t+B)}\right)$ 
& $\frac{\ln{\left(\frac{\left(B + x\right)^{2}}{a \cosh^{2}{\left(A + t \right)}} \right)}}{\lambda}$ 
& $\equiv$ \\[6pt]

$\frac{\partial^2 u}{\partial t^2} - a\exp(\lambda u)\frac{\partial^2 u}{\partial x^2} = 0$ 
& $\ln\left(\frac{\left(\exp\left(x\right)+\exp\left(-x\right)\right)^2}{4}\right)$ 
& $\frac{1}{\lambda}\ln\left(\frac{\left(A\exp\left(x\right)+B\exp\left(-x\right)\right)^2}{4aAB\left(t+C\right)^2}\right)$ 
& $\frac{\ln{\left(\frac{\left(B \exp(- x) + C \exp(x)\right)^{2}}{4 B C a \left(A + t\right)^{2}} \right)}}{\lambda}$ 
& $\equiv$ \\[6pt]

\end{longtable}
}
\end{landscape}

\restoregeometry

	\subsection{Case study}	
	\begin{figure*}[t]
		\centering
		\includegraphics[width=1.0 \textwidth]{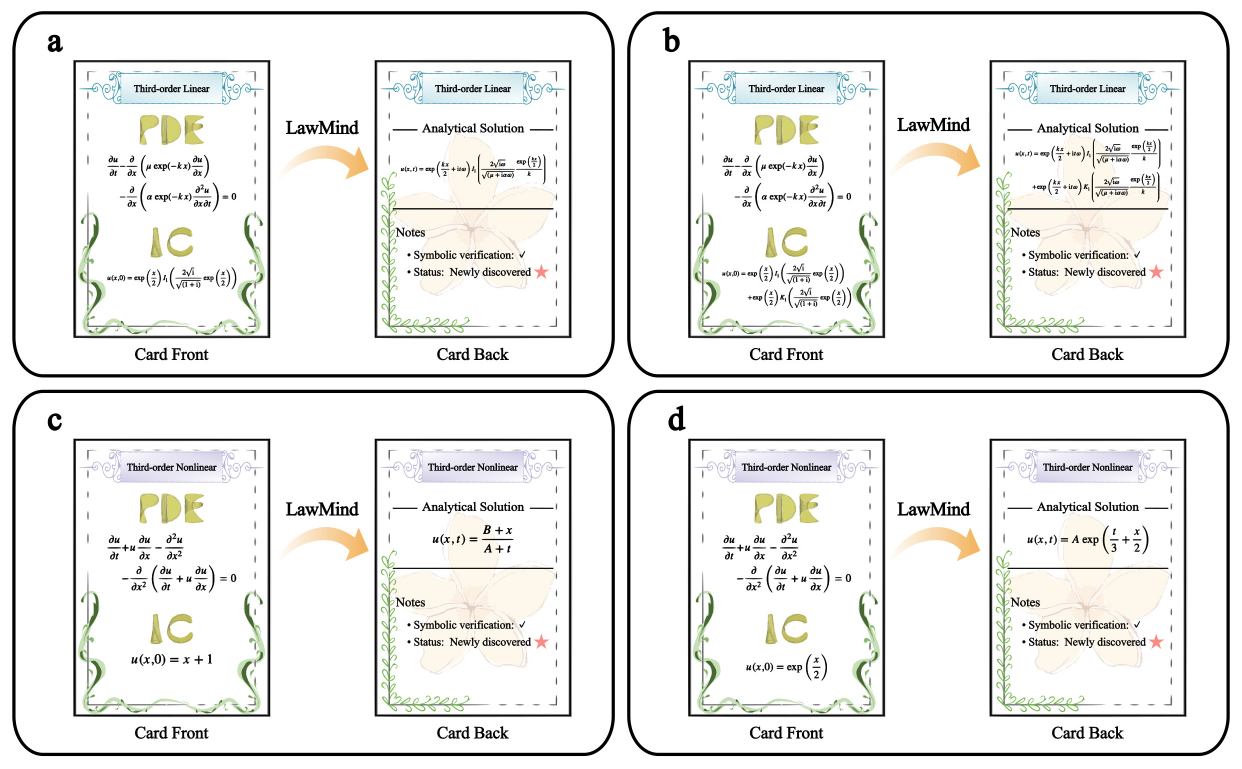}
		\caption{\textbf{Closed-form solutions beyond classical handbooks}. 
			\textbf{a}, \textbf{b}, Two newly discovered analytical solutions to a third-order linear PDE under distinct initial conditions. 
			\textbf{c}, \textbf{d}, Two newly discovered analytical solutions to a third-order nonlinear PDE under distinct initial conditions. For each case, the card front presents the governing PDE and initial condition, and the card back presents the closed-form solution recovered by LawMind.}
		\label{fig:result3}
	\end{figure*}
	
	The space of PDEs with documented closed-form solutions remains limited. Classical handbooks systematically cover equations of standard form, yet even minor structural departures, such as the introduction of spatially varying coefficients, mixed partial derivatives, or nonlinear terms, often leave the resulting equations without documented closed-form solutions. This motivates examining whether LawMind produces closed-form solutions for equations beyond the coverage of existing reference tables.
	
	We consider two third-order PDEs not covered by the benchmark in \cref{ss:Benchmark Design and Recovery}. One is a third-order linear PDE representing a special viscoelastic fluid with spatially varying viscosity coefficients, relevant to the vibrational behavior of materials with memory effects. Its structure falls outside the standard parabolic form covered by classical handbooks. The other is a third-order nonlinear PDE describing a generalized viscous Burgers model, where the material derivative is regularized by a Helmholtz operator. This structural extension of the classical equation is relevant to the extreme behavior of hydrodynamic waves with extra dispersion, and likewise lacks documented closed-form solutions. Each equation is evaluated under two distinct initial conditions, yielding four discovery tasks in total, as illustrated in \cref{fig:result3}.
	
	LawMind successfully discovered closed-form solutions for all four cases, each satisfying the governing equation and initial condition under symbolic identity verification. The governing equations, initial conditions, and discovered solutions are documented in \cref{tab:newly_discovered}. To the best of our knowledge, these solutions have not been previously documented in classical handbooks or the broader analytical literature.
	
	For equations that lack established solution formulas, LawMind discovers closed-form solutions directly from the governing equation and initial condition, without relying on predefined solution templates or transformation techniques. These results demonstrate that law-driven symbolic discovery produces closed-form solutions for equations that lie beyond the coverage of classical solution tables, where established analytical procedures do not readily apply. For researchers working with such equations, closed-form solutions provide direct access to parameter dependencies and scaling behaviors that numerical approaches cannot readily reveal.
	
	\begin{table}
	\centering
	\tiny
	\renewcommand{\arraystretch}{2.5}
	\caption{\normalsize{Newly Discovered Analytical Solutions}}
	\label{tab:newly_discovered}
	\begin{tabular}{lll}
		\toprule
		{\normalsize\textbf{PDE}} & {\normalsize\textbf{Initial Condition}} & {\normalsize\textbf{Discovered Analytical Solution}} \\
		\midrule
		\multicolumn{3}{l}{\scriptsize\textbf{Third-order Linear PDEs}} \\
		\midrule
		
		\multirow{2}[12]{*}{$\splitfrac{\frac{\partial u}{\partial t} - \frac{\partial }{\partial x}\left( \mu \exp(-kx) \frac{\partial u}{\partial x}\right)}{ + \frac{\partial }{\partial x}\left(\alpha \exp(-kx) \frac{\partial^2 u}{\partial x \partial t} \right)  = 0}$}
		& $\exp\left(\frac{x}{2}\right)I_1\left(\frac{2\sqrt{\mathrm{i}} }{\sqrt{(1 +\mathrm{i})}}\exp\left({\frac{x}{2}}\right)\right)$
		& $\exp\left(\frac{k x}{2}+\mathrm{i} t \omega\right)
		I_1\left(\frac{2\sqrt{\mathrm{i} \omega} }{\sqrt{(\mu +\mathrm{i} \alpha \omega)}}\frac{\exp\left({\frac{k x}{2}}\right)}{k}\right)$ \\[15pt]
		
		& $\splitfrac{\exp\left(\frac{x}{2}\right) I_1\left(\frac{2\sqrt{\mathrm{i}} }{\sqrt{(1 +\mathrm{i})}}\exp\left({\frac{x}{2}}\right)\right)}
		{ + \exp\left(\frac{x}{2}\right) K_1\left(\frac{2\sqrt{\mathrm{i}} }{\sqrt{(1 +\mathrm{i})}}\exp\left({\frac{x}{2}}\right)\right)}$
		& $\splitfrac{\exp\left(\frac{k x}{2}+\mathrm{i} t \omega\right)
			I_1\left(\frac{2\sqrt{\mathrm{i} \omega} }{\sqrt{(\mu +\mathrm{i} \alpha \omega)}}\frac{\exp\left({\frac{k x}{2}}\right)}{k}\right)}
		{+ \exp\left(\frac{k x}{2}+\mathrm{i} t \omega\right)K_1\left(\frac{2\sqrt{\mathrm{i} \omega} }{\sqrt{(\mu +\mathrm{i} \alpha \omega)}}\frac{\exp\left({\frac{k x}{2}}\right)}{k}\right)}$ \\
		
		\midrule
		\multicolumn{3}{l}{\scriptsize\textbf{Third-order Nonlinear PDEs}} \\
		\midrule
		
		\multirow{2}[2]{*}{$\splitfrac{\frac{\partial u}{\partial t} + u\frac{\partial u}{\partial x} - \frac{\partial^2 u}{\partial x^2}}{ - \frac{\partial}{\partial x^2}\left( \frac{\partial u}{\partial t} + u\frac{\partial u}{\partial x} \right)  = 0}$}
		& $x+1$
		& $\frac{B+x}{A+t}$ \\[6pt]
		
		& $\exp\left(\frac{x}{2}\right)$
		& $A\exp\left(\frac{t}{3}+\frac{x}{2}\right)$ \\
		
		\bottomrule
	\end{tabular}
\end{table}
	
	\section{Discussion}\label{sec:conclusion}
	
	This work demonstrates that closed-form analytical solutions to partial differential equations are discoverable directly from governing equations through law-driven symbolic search, without reliance on numerical data, supervised training, or predefined solution templates. Validated on a benchmark of 100 PDEs drawn from authoritative reference handbooks and spanning five canonical categories, LawMind successfully recovered parametric closed-form solutions for all problems. Each solution was confirmed to satisfy the governing equation and initial condition under symbolic identity verification. Beyond benchmark recovery, LawMind discovered closed-form solutions for equations that fall outside the coverage of classical solution tables, including third-order linear and nonlinear PDEs for which no analytical solutions have been previously documented. These results establish law-driven symbolic discovery as a principled approach to constructing exact analytical solutions for equations that classical solution procedures do not systematically address.
	
	To situate this contribution, it is useful to consider how LawMind relates to existing approaches. Numerical and data-driven methods, such as finite element solvers and physics-informed neural networks, produce approximate solutions that depend on discretization or training data and do not yield closed-form expressions. LawMind addresses a fundamentally different objective, namely the construction of exact parametric closed-form solutions, and therefore operates in a distinct problem space from these methods.
	
	Classical analytical methods, such as separation of variables, integral transforms, and reduction to ordinary differential equations, represent the most direct precedent for LawMind, as both aim at exact closed-form solutions. These procedures are effective for equations of standard form but rely on recognizing structural patterns in the governing equation and selecting appropriate transformations based on prior mathematical knowledge. Their applicability is therefore tied to the availability of established solution strategies for a given equation class. LawMind takes a different path by conducting symbolic search directly on the governing equation and initial condition, without presupposing a transformation strategy or solution template. This distinction explains why LawMind produces closed-form solutions for equations that fall outside the systematic coverage of classical handbooks, where traditional analytical procedures do not readily apply.
	
	The current framework carries two limitations that define the boundary of its applicability. The first concerns the scalability of the search procedure. The scale of the search space is determined jointly by the number of symbolic variables and parameters in the target solution, the complexity of the initial condition, and the size of the terminal set. As any of these factors grows, the number of candidate expressions increases combinatorially, such that solutions of high structural complexity may exceed the practical reach of the current search procedure. Within the scope of the benchmark evaluated in this work, the search remains computationally tractable, but the combinatorial growth of candidate expressions with increasing solution complexity presents a fundamental challenge for scaling to more intricate equation families.
	
	Additionally, the present framework is designed for PDEs with a time variable, where an initial condition provides the structural anchor for solution construction. This dependence arises because the discovery process uses the initial condition as the starting point from which symbolic expressions are assembled and evaluated. Extension to equations governed solely by boundary conditions, without a time evolution structure, remains outside the scope of the current formulation.
	
	Several directions present natural extensions of the present work. The most immediate concerns search efficiency, where incorporating learning-driven prioritization into the symbolic search procedure could reduce the combinatorial growth of candidate expressions and extend the practical reach of the framework to solutions of greater structural complexity. A second direction involves broadening the class of equations that the framework addresses. The current formulation relies on a time evolution structure to anchor solution construction. Developing analogous mechanisms for equations governed solely by boundary conditions represents a natural direction for expanding the scope of law-driven symbolic discovery. Third, the framework extends conceptually to systems of coupled PDEs, where the verification criterion generalizes to requiring simultaneous satisfaction of multiple governing equations, with the primary challenge lying in controlling the scale of the joint search space across multiple unknown functions. Finally, exact parametric closed-form solutions discovered through this paradigm provide direct analytical tools for scientific domains such as fluid mechanics, heat transfer, and wave dynamics. In these domains, closed-form expressions support physical interpretation and enable exact computation in ways that numerical approximations do not.
	
	The discovery of closed-form solutions to partial differential equations has historically depended on mathematical intuition and the recognition of structural patterns amenable to known analytical techniques. LawMind demonstrates that this process is amenable to systematic symbolic search, enabling closed-form solutions to be constructed directly from governing equations without reliance on prior solution strategies or human-guided transformations. By establishing a repeatable and operationalizable pathway to exact analytical solutions, law-driven symbolic discovery shifts the construction of closed-form solutions from an insight-dependent endeavor to a structured computational process. These results demonstrate that the set of known closed-form solutions to partial differential equations is not exhaustive, and that systematic symbolic search provides a principled means of extending it.
	
	\section{Methods}\label{sec:methods}
	
	LawMind is a law-driven symbolic discovery framework that recovers closed-form analytical solutions to PDEs directly from governing equations and associated conditions, without relying on numerical data or external supervision. The following sections formulate the problem and describe the four components of the framework, including structural assembly, evolutionary expansion, hierarchical search space construction, and law-driven symbolic evaluation. These components operate within a closed loop, as summarized in \cref{fig:result1}.
	
	\subsection{Problem formulation}\label{ss:Problem Formulation}
	
	Let the governing equation and its associated conditions be denoted as $(\mathcal{N}, C)$, where $\mathcal{N}[u(x,t);\bm{\lambda}] = 0$ is a differential operator acting on an unknown field $u(x,t)$ with physical coefficients $\bm{\lambda}$, and $C$ collectively denotes the imposed initial, boundary, or source conditions. The objective of LawMind is to recover a closed-form analytical expression $\hat{u}(x,t)$ such that
	$$
	\mathcal{N}[\hat{u}(x,t);\bm{\lambda}] = 0
	$$
	holds identically in the symbolic sense, and $\hat{u}$ is consistent with all conditions in $C$.
	
	The solution $\hat{u}$ is represented as a symbolic expression tree, whose leaf nodes correspond to independent variables and constants, and whose internal nodes correspond to mathematical operators. The search is conducted entirely in the symbolic domain. Satisfaction of the governing equation is verified through exact symbolic computation of the differential operator rather than pointwise numerical evaluation, ensuring that discovered solutions are analytically exact. This formulation requires no discretization of the governing equation and no numerical data, and is applicable to a broad class of PDEs including those that are nonlinear or high-order.
	
	\subsection{Structural assembly}\label{ss:Structural Assembly}
	
	In the search phase, candidate analytical solutions are represented as symbolic expression trees. Leaf nodes are drawn from a terminal set $\mathcal{T}$, and internal nodes are drawn from a function set $\mathcal{F}$. The position set $\mathcal{P}$ comprises all nodes of the current tree, each serving as a valid site for subtree insertion. An insertion at node $p \in \mathcal{P}$ introduces a new internal node from $\mathcal{F}$, with the original node and a new leaf from $\mathcal{T}$ assigned as its two children. The left-right assignment of the two children is treated as an additional binary choice.
	
	At each stage $k$, LawMind assembles candidate subtrees from the available terminal set $\mathcal{T}^{(k)}$, function set $\mathcal{F}$, and position set $\mathcal{P}^{(k)}$. Each candidate subtree is formed by combining an element from $\mathcal{P}^{(k)}$, an operato from $\mathcal{F}$, and a terminal expression involving the newly activated variable $v_k$ from $\mathcal{T}^{(k)}$. The set of all such combinations constitutes the structural candidate pool at stage $k$, and each candidate subtree represents a potential expansion of the current expression at a designated insertion site.
	
	\subsection{Evolutionary expansion}\label{ss:Evolutionary Expansion}
	
	Given the structural candidate pool assembled at stage $k$, LawMind expands the current expression by inserting a candidate subtree from the pool at its designated position within the current expression tree. At the first stage, the current expression is initialized from the initial condition $u(x,0) = g(x)$, providing a starting point grounded in the physical specification of the problem. At each subsequent stage $k$, the current expression is the analytical solution identified at stage $k-1$, and expansion proceeds from this validated expression.
	
	The search terminates as soon as a candidate expression is found to satisfy $(\mathcal{N}, \mathcal{C})$, at which point the process advances to stage $k+1$ without further enumeration. In principle, the search is exhaustive given sufficient computational resources. In practice, a computational budget is imposed to ensure tractability.
	
	\subsection{Hierarchical search space construction}\label{ss:Hierarchical Search Space Construction}
	
	A central challenge in symbolic search is the combinatorial growth of the search space as the number of variables increases. Without hierarchical activation, all variables are introduced simultaneously into $\mathcal{T}$, and the same $|\mathcal{P}|$ positions must accommodate repeated insertions to construct expressions involving all variables. Allowing up to $M$ insertions per position, the search space grows as
	\begin{equation*}
		|\mathcal{S}| = O\left((2|\mathcal{F}||\mathcal{T}|)^{|\mathcal{P}| \times M}\right),
	\end{equation*}
	where $M \geq n$ is required to incorporate all $n$ variables, and the complexity grows exponentially with $n$.
	
	To address this, LawMind activates variables into $\mathcal{T}$ incrementally across successive stages. At each stage $k$, the terminal set $\mathcal{T}^{(k)}$ is constructed from the newly activated variable $v_k$ and its admissible variants, reflecting the variables that remain to be incorporated into the current expression. Since each position admits at most one insertion per stage, the search space at stage $k$ is
	\begin{equation*}
		|\mathcal{S}^{(k)}| = O\left((2|\mathcal{F}||\mathcal{T}^{(k)}|)^{|\mathcal{P}^{(k)}|}\right),
	\end{equation*}
	where the factor $2|\mathcal{F}||\mathcal{T}^{(k)}|$ accounts for the choice of operator, new leaf, and left-right assignment at each position. Upon discovery of a valid analytical solution at stage $k$, this solution is carried forward to stage $k+1$ as the current expression, and the 
	process continues until all variables have been activated and a complete analytical solution satisfying $(\mathcal{N}, \mathcal{C})$ is identified. The total complexity across all $n$ stages is
	\begin{equation*}
		|\mathcal{S}_{\text{hier}}| = O\left(\sum_{k=1}^{n}
		(2|\mathcal{F}||\mathcal{T}^{(k)}|)^{|\mathcal{P}^{(k)}|}\right).
	\end{equation*}
	Since $|\mathcal{T}^{(k)}| \leq |\mathcal{T}|$ and $|\mathcal{P}^{(k)}| \ll |\mathcal{P}| \times M$, each stage-wise term satisfies $(2|\mathcal{F}||\mathcal{T}^{(k)}|)^{|\mathcal{P}^{(k)}|} \ll 
	(2|\mathcal{F}||\mathcal{T}|)^{|\mathcal{P}| \times M}$, and consequently 
	$|\mathcal{S}_{\text{hier}}| \ll |\mathcal{S}|$. Hierarchical activation thus decomposes the exponential growth of the flat search into a sum of stage-wise terms, each governed by the structure of the current expression rather than the full variable set, rendering the symbolic search tractable at each stage while enabling the progressive construction of analytical 
	solutions of increasing complexity.
	
	\subsection{Law-driven symbolic evaluation}\label{ss:Law-Driven Symbolic Evaluation}
	
	Each candidate expression generated during evolutionary expansion is evaluated through exact symbolic verification against the governing law. This evaluation proceeds in two sequential steps.
	
	First, the candidate expression $\hat{u}$ is substituted into the governing equation $\mathcal{N}[\hat{u};\mathbf{\lambda}] = 0$, and the residual is computed symbolically with all variables and parameters treated as symbolic quantities. The candidate passes the PDE residual check if and only if the residual vanishes identically, confirming that the governing equation is satisfied for arbitrary values of all variables and parameters rather than approximately or at discrete sample points.
	
	Second, if the PDE residual check is passed, the candidate is further verified for consistency with the associated conditions $C$. The initial condition check is performed by substituting $t=0$ into the candidate expression and setting any symbolic parameters to their reference values, and the candidate passes if and only if the result reduces identically to $u(x,0)$.
	
	A candidate expression is accepted as a valid analytical solution at stage $k$ if and only if it passes both checks. If either check fails, the candidate is rejected and the search continues over the remaining structural candidates at stage $k$. Fitness is assigned in a strict binary manner, reflecting the requirement that discovered solutions are mathematically exact rather than approximate.
	
	Due to the progressive variable activation mechanism, different activation sequences may yield parametric representations of the same solution family that differ in form but are mathematically equivalent. Consistency with reference solutions is therefore assessed by verifying mathematical equivalence rather than requiring identical parametric forms.
	
	This law-driven evaluation eliminates the need for numerical simulation or training data, and ensures that every solution returned by LawMind exactly satisfies both the governing equation and the associated conditions.	
	
	\bibliography{sn-bibliography}
	
\end{document}